# Comment: Struggles with Survey Weighting and Regression Modeling

**Danny Pfeffermann**

This is an intriguing paper that raises important questions, and I feel privileged for being invited to discuss it. The paper deals with a very basic problem of sample surveys: how to weight the survey data in order to estimate finite population quantities of interest like means, differences of means or regression coefficients.

The paper focuses for the most part on the common estimator of a population mean, $\bar{y}_w = \sum_{i=1}^{n} w_i y_i / \sum_{i=1}^{n} w_i$, and discusses different approaches to constructing the weights by use of linear regression models. These models vary in terms of the number and nature of the regressors in the model and in the assumptions regarding the regression coefficients, whether fixed or random with prespecified distributions. The idea behind regression weighting is to include in the regression model all the variables and interactions that are related to the outcome values and affect the sample selection and the response probabilities, such that the sampling and response mechanisms are ignorable in the sense that the model fitted to the observed data is the same as the population model before sampling. Assuming that all the important regressors affecting the sample selection and response are discrete, the set of all possible combinations of categories of these variables defines poststratification cells, which in turn define the dummy independent variables in the regression model. The target population parameter of interest can be written then as $\theta = \sum_{j=1}^{J} N_j \theta_j / \sum_{j=1}^{J} N_j$, where $\theta_j$ is the parameter for cell $j$ (say the true cell mean, $\bar{Y}_j$), $N_j$ is the cell size and $J$ is the number of cells. The regression estimator has the general form $\hat{\theta}^{PS} = \sum_{j=1}^{J} N_j \hat{\theta}_j / \sum_{j=1}^{J} N_j$. For example, the estimator of the population mean is $\hat{\bar{Y}}^{PS} = \sum_{j=1}^{J} N_j \bar{y}_j / \sum_{j=1}^{J} N_j$, where $\bar{y}_j$ is the sample mean in cell $j$.

*Danny Pfeffermann is Professor, Department of Statistics, Hebrew University of Jerusalem, Jerusalem 91905, Israel and Professor, Southampton Statistical Sciences Research Institute, University of Southampton, Southampton SO17 1BJ, United Kingdom e-mail: msdanny@huji.ac.il.*

The discussion that follows is divided into two parts. In the first part I comment on the proposed weighting approach and some of the statements made in the article. In the second part I consider another approach for analyzing survey data that are subject to unequal sample selection probabilities and nonresponse, and compare it to the approach taken in this paper.

## 1. REMARKS ON THE PAPER

• The first obvious remark, that is also made already in the Abstract, is that the number of poststratification cells can be extremely large, with inevitably very small or no samples in many of the cells. Having small or no samples in some or even most of the cells is theoretically not a problem under the model with random regression coefficients considered later, but it is not clear what should be done in such a case under the standard regression model with fixed coefficients. Note in particular the problems arising if the zero sample sizes are due to nonresponse. Deleting these cells from the regression model may violate the sample ignorability assumption. It is stated in Section 3.1 that it is not required to include in the model all the cells, but this raises the question of which cells to exclude and based on what criteria. It may imply also including different cells (interactions) for different outcome variables of interest.

• It is assumed that the cell sizes are known. This could be a strong assumption in a large-scale survey with many small cells. Also, it is often the case that the cell sizes are known to the person drawing the sample, but not necessarily to the person analyzing the data, who has limited access to the original files due to confidentiality restrictions or other reasons. The argument that the cell sizes can be estimated using iterative proportional fitting or other related procedures is well taken, but this raises questions of the effect of using estimated sizes on the performance of the estimators and how to estimate the variances, accounting for this source of variability.

• A third problem and in a way the most difficult one to handle is the implicit assumption that the analyst knows all the variables affecting the sample selection and nonresponse. Here again a distinction should be made between the person drawing the sample who should at least know all the variables that affect the sample selection, and the person analyzing the data who may not even have that information. When it comes to nonresponse, both persons can only hypothesize which variables explain the nonresponse. I should add to this that the paper implicitly assumes that the missing data are missing at random (MAR), which of course may not be the case in practice. The alternative approach described later overcomes in principle these problems but it requires modeling the sample inclusion probabilities as a function of the observed data.

• It is mentioned that computing the variances of weighted estimators may not be trivial, because the weights are generally random variables that depend on the data. I can see that weighting cells that account for nonresponse are "data driven," but for given cells, the computation of the variances should not be complicated, even though the response probabilities are only estimated. Thus, a distinction should be made between conditional and unconditional variances. A more crucial distinction, however, is between variances and mean square errors, because as already implied by my previous comment, the main issue is whether the cells are defined correctly and the nonresponse is indeed MAR.

• The paper proposes a two-step procedure for estimating the regression of $y$ on $z$. The first step consists of regressing $y$ on $z$ and $X$ and interactions between them, where $X$ represents the variables affecting the sample inclusion probabilities; the second step consists of regressing $X$ on $z$ in order to obtain the regression of $y$ on $z$ alone [(4) in the paper]. I have no problem with this approach, but as the paper repeatedly emphasizes the regression (averaging) in the second step must be adjusted for the population distribution of $X$. If this distribution is unknown, which may well be the case in practice, one is bound to use some sort of weighting in the back door. Thus, an alternative "weighted regression" procedure favored by survey analysts is to regress $y$ on $z$ alone, but use weighted regression with the weights defined by the inverse of the sample inclusion probabilities. Consider the example in the paper of regressing log earning against ethnicity (white/nonwhite) in order to estimate the difference $E(y|\text{white} = 1) - E(y|\text{white} = 0)$. Suppose that the survey oversamples males. It is argued that the model should include in this case as additional regressors "gender" and the interaction between white and male, and then obtain the regression of log earning on ethnicity by applying the second step described above. This model accounts for possible differences between the effects of the two genders on the log earning for a given ethnicity, and is thus the "correct model," irrespective of the sample inclusion probabilities. Application of the weighted regression approach to the example consists in this case of regressing $y$ against $z$ (defined by two dummy variables representing "white" and "nonwhite") and weighting each sample value by the inverse of the sample inclusion probability. Denoting the sample of "white" by $S_1$ and the sample of "nonwhite" by $S_2$, the resulting estimator is $(\sum_{i \in S_1} w_i y_{i1} / \sum_{i \in S_1} w_i) - (\sum_{i \in S_2} w_i y_{i1} / \sum_{i \in S_2} w_i)$. Clearly, if the model with the gender variable and the interaction term is the correct model, the model without them is the wrong model and weighting the sample observations does not correct the model. However, as long as the weights are estimated appropriately (accounting for the sample selection and response probabilities), the use of this procedure yields a consistent estimator for the difference of interest. I believe that many analysts would use weighted regression even when fitting the "correct model," so as to protect against other possible model misspecifications.

• It is mentioned in Section 3.1 that the full poststratification estimator of the population mean, $\hat{\theta}^{PS} = \sum_{j=1}^{J} N_j \bar{y}_j / \sum_{j=1}^{J} N_j$, can be viewed as a classical regression estimator by including indicators for all the poststratification cells. How are the sizes $N_j$ captured by the regression model? Is it not a weighted regression estimator?

• It is stated that weighted regression is not flexible and that it is not clear how to apply the weights. I do not think that this is correct. The use of pseudo likelihood methods, for example (see the discussion and references in Pfeffermann, 1993), is well established and very common. See also below for a model-based justification for weighted regression. The use or nonuse of the weights has nothing to do with the use of models for small area estimation problems, as seems to be suggested in Section 4.

• As pointed out in the paper, the use of hierarchical models implies different sets of weights for different outcome variables. Statistical bureaus do not like this and usually insist on a single weight for a given sample unit, even at the risk of loss of efficiency. To highlight this problem a bit further, suppose that one is interested in three variables, $y_1$, $y_2$ and $y_3 = g(y_1, y_2)$ for some function $g$. Say $y_1$ is total earnings in a given month, $y_2$ is the number of hours worked and $y_3$ is the mean earning per hour. Fitting a hierarchical model to each of the three variables separately would imply three different sets of weights, which some would argue does not make sense in this case, beyond not being practical.

• The use of the normal model with independent random effects for the $J$ cell means does not seem appropriate if the cells are defined by interactions of the regressors that account for the sample selection and nonresponse. Some of these cells are "close" to each other, say the cells defined by given categories of gender, ethnicity and level of education, and adjacent categories of age, and other cells are very apart. Thus, it is more appropriate in this case to fit a model with spatial correlations between the random effects that reflect the distance between the corresponding cells. The computation of the weights under the model is neat. Note that with many cells and very small sample sizes within the cells, the cell predictor $\hat{\theta}_k$ in (10) will often be close to the synthetic estimator $\hat{\mu}$ in (11), which is then also approximately the estimator of the population mean. As a result, the weight will be approximately constant.

## 2. ALTERNATIVE APPROACHES

As discussed above, a major problem with the application of the approach proposed in this article is that it requires knowledge of all the important variables affecting the sample selection or nonresponse (the $X$ variables). As argued by Alexander (1987), "no model will include all the relevant variables and few analysts will wish to include in the model all the geographic and operational variables which determine sampling rates. The theoretical and empirical tasks of fitting and validating such models seem formidable for many surveys."

One way to deal with this problem, considered by Rubin (1985), is to use the vector of sample inclusion probabilities as a surrogate for the variables in $X$, but as further discussed in Smith (1988), this approach is not always valid and in the case of nonresponse, the true inclusion probabilities are unknown and need to be estimated. Skinner (1994) models the outcomes in the sample as a function of the model covariates and the sampling weights, and the sampling weights in the sample as a function of the model covariates, and shows how to obtain the model for the outcomes in the population from these two models.

In what follows I outline briefly the basic ideas of another approach for estimating population models and predicting finite population quantities. This approach models the sample data and bases the inference on the sample model. See the references below for more details with examples and applications. I consider for convenience single stage sampling and assume that the sample selection and response are independent between the sampling units. As before, denote by $y$ the outcome variable and suppose first that one is interested in identifying and estimating the population model $f_p(y|z)$, where $z$ is a set of covariates. Following Pfeffermann, Krieger and Rinott (1998), the sample model is defined as

$$
\begin{aligned}
f_s(y_i|z_i) &\overset{\text{def}}{=} f(y_i|z_i, i \in s) \\
&= \frac{\Pr(i \in s|y_i, z_i) f_p(y_i|z_i)}{\Pr(i \in s|z_i)} \\
&= \frac{E_p(\pi_i|y_i, z_i) f_p(y_i|z_i)}{E_p(\pi_i|z_i)},
\end{aligned}
$$

where $\pi_i = \Pr(i \in s)$ is the sample inclusion probability (probability to be selected and respond).

REMARK 1. By (1), the sample model is the same as the population model if $\Pr(i \in s|y_i, z_i) = \Pr(i \in s|z_i)$ $\forall y_i$, in which case the sampling process is ignorable.

REMARK 2. $\Pr(i \in s|y_i, z_i)$ is generally not the same as $\pi_i$, which may depend on the variables in $X$ and possibly also on the $y$-values in the case of NMAR nonresponse. However, the use of the sample model only requires modeling $\Pr(i \in s|y_i, z_i)$ or $E_p(\pi_i|y_i, z_i)$, thus circumventing the need to know the variables $X$ and incorporate them in the model. Note that the sample model resulting from modeling the sample inclusion probabilities can be tested using standard goodness-of-fit test statistics, since the sample model refers to the sample data.

The following relationship between the population model and the sample model is established in Pfeffermann and Sverchkov (1999), where $w_i = 1/\pi_i$ and

$E_s(\cdot)$ is the expectation under the sample model:

$$(2) \qquad f_p(y_i|z_i) = \frac{E_s(w_i|y_i, z_i) f_s(y_i|z_i)}{E_s(w_i|z_i)}.$$

Thus, one can identify and estimate the population model by fitting the sample model to the sample data and estimating the expectations $E_s(w_i|y_i, z_i)$, again using the sample data. Clearly, both the sample model and the expectations $E_s(w_i|y_i, z_i)$ depend in general on unknown parameters. Pfeffermann and Sverchkov (2003) discuss alternative approaches of estimating these parameters, with examples. Note in this respect that if the outcomes are independent under the population model, they are also "asymptotically independent" under the sample model when increasing the population size but holding the sample size fixed. See Pfeffermann, Krieger and Rinott (1998) for details.

REMARK 3. For likelihood- or Bayesian-based inference, one can employ the "full likelihood" of the sample data and the sample membership indicators,

$$(3) \qquad \begin{aligned} f(s, y_s|z_s, z_{\bar{s}}) &= \prod_{i \in s} \Pr(i \in s|y_i, z_i) f_p(y_i|z_i) \\ &\quad \cdot \prod_{j \notin s} [1 - \Pr(j \in s|z_j)], \end{aligned}$$

where $\Pr(j \in s|z_j) = \int \Pr(j \in s|y_j, z_j) f_p(y_j|z_j) \, dy_j$ is the propensity score for unit $j$; see, for example, Gelman et al. (2004) and Little (2004). The use of (3) has the advantage of employing the information on the sample selection probabilities for units outside the sample, but it requires knowledge of the covariates for every unit in the population, unlike the use of the sample likelihood that is based on the sample model. Modeling the joint distribution of the covariates and integrating them out of the likelihood is often too complicated.

REMARK 4. I mentioned before that the use of weighted regression can be justified theoretically. Suppose that the population model is $y_i = z_i'\beta + \varepsilon_i$; $E_p(\varepsilon_i|z_i) = 0$, $E_p(\varepsilon_i^2|z_i) = \sigma_\varepsilon^2$. By (2),

$$(4) \qquad \begin{aligned} \beta &= \arg\min_{\tilde{\beta}} E_p(y_i - z_i'\tilde{\beta})^2 \\ &= \arg\min_{\tilde{\beta}} E_s\left[\frac{w_i(y_i - z_i'\tilde{\beta})^2}{E_s(w_i)}\right] \\ &= \arg\min_{\tilde{\beta}} E_s[w_i(y_i - z_i'\tilde{\beta})^2], \end{aligned}$$

noting that $E_s(w_i) = [N/E(n)] = const.$ Replacing the sample expectation in the right-hand side of (4) by the sample mean yields the weighted regression estimator $b_w = [\sum_{i \in s} w_i z_i z_i']^{-1} \sum_{i \in s} w_i z_i y_i$ as the optimal (least squares) solution.

REMARK 5. By conditioning on $z_i$ and hence minimizing $E_s[\frac{w_i(y_i - z_i'\tilde{\beta})^2}{E_s(w_i|z_i)}|z_i]$, one obtains the estimator $b_q = [\sum_{i \in s} q_i z_i z_i']^{-1} \sum_{i \in s} q_i z_i y_i$, where $q_i = w_i/E_s(w_i|z_i)$. The weights $\{q_i\}$ account for the net sampling effects on the conditional target distribution $f_p(y_i|z_i)$, and the estimator $b_q$ is therefore less variable than $b_w$. See Pfeffermann and Sverchkov (1999) for further discussion and empirical comparisons between the two estimators.

How can the sample model be used for estimating finite population totals or means? For this we need to define the sample-complement model,

$$(5) \qquad \begin{aligned} f_c(y_i|z_i) &\overset{\text{def}}{=} f(y_i|z_i, i \notin s) \\ &= \frac{\Pr(i \notin s|y_i, z_i) f_p(y_i|z_i)}{\Pr(i \notin s|z_i)} \\ &= \cdots = \frac{E_s[(w_i - 1)|y_i, z_i] f_s(y_i|z_i)}{E_s[(w_i - 1)|z_i]}, \end{aligned}$$

with the last equality shown in Sverchkov and Pfeffermann (2004). Note that the sample-complement model is again a function of the sample model and the expectation $E_s(w_i|z_i)$, and thus can be estimated from the sample data. The optimal predictor of the population total under a quadratic loss function is,

$$(6) \qquad \begin{aligned} \hat{Y} &= \sum_{i \in s} y_i + \sum_{j \notin s} E(y_j|z_j, j \notin s) \\ &= \sum_{i \in s} y_i + \sum_{j \notin s} E_c(y_j|z_j) \\ &= \sum_{i \in s} y_i + \sum_{j \notin s} \frac{E_s[(w_j - 1)y_j|z_j]}{E_s[(w_j - 1)|z_j]}. \end{aligned}$$

The last equality follows from (5), with the sample expectations in the numerator and the denominator either being modeled based on the sample data or simply estimated by the corresponding sample means by application of the method of moments. As shown in Sverchkov and Pfeffermann (2004), familiar estimators of finite population means such as the estimator $\bar{y}_w = \sum_{i=1}^n w_i y_i / \sum_{i=1}^n w_i$ studied in the present paper are obtained as special cases of this theory by specifying appropriate population or sample models. Pfeffermann and Sverchkov (2007) use the sample and sample-complement models for small area estimation under informative sampling of areas and within the areas.

To summarize, the alternative approach outlined above has the advantage of not requiring incorporating in the model the variables affecting the sample selection and response, unless they are part of the covariates that define the target model of interest. It can be applied also in situations where the response process is NMAR. However, it requires modeling the expectation $E_s(w_i|y_i, z_i)$, which may not be easy in the presence of nonresponse. On the other hand, as mentioned before, the resulting sample model can be tested using classical goodness-of-fit statistics, since the sample model refers to the sample data.